\newcommand*{\TechReport}{}%
\newcommand*{\AcmFormat}{}%

\ifdefined\TechReport
\documentclass[10pt,sigconf,letterpaper]{acmart}
\else
\documentclass[10pt,sigconf,letterpaper,anonymous]{acmart}
\fi

\usepackage{times}  
\ifdefined\AcmFormat
\else
\usepackage[hidelinks]{hyperref}

\usepackage{graphicx}
\usepackage{cite}
\usepackage[cmex10]{amsmath}
\usepackage{amssymb}
\usepackage{comment}
\usepackage{float}
\usepackage{subcaption}
\usepackage{url}
\usepackage{amsfonts}
\usepackage{amsthm}

\fi
\usepackage{balance}

\ifdefined\AcmFormat
\else
\fi

\usepackage{array}
\usepackage{subcaption}
\usepackage{enumitem}

\newlength{\grafflecm}
\ifdefined\AcmFormat

\renewcommand\footnotetextcopyrightpermission[1]{} 
\fi
\begin{document}

\makeatletter
\ifdefined\AcmFormat
\renewcommand\@formatdoi[1]{\ignorespaces}
\makeatother
\fi
\pagestyle{plain}

\date{}

\title{
Internet Performance in the 2022 Conflict in Ukraine: An Asymmetric Analysis
}

\ifdefined\AcmFormat

\author{Tal Mizrahi}
\affiliation{%
  \institution{Technion --- Israel Institute of Technology}
}

\author{Jose Yallouz}
\affiliation{%
  \institution{Technion --- Israel Institute of Technology}
}

\else

\fi

\thispagestyle{empty}

\ifdefined\AcmFormat
\begin{abstract}
On 24 February 2022 Russia invaded Ukraine, starting one of the largest military conflicts in Europe in recent years. In this paper we present preliminary findings about the impact of the conflict on the Internet performance in Ukraine and in Russia, introducing an ironically asymmetric picture: the Internet performance in Ukraine has significantly degraded, while the performance in Russia has improved. 
\end{abstract}

\maketitle

\section{Introduction}
\label{IntroSec}
The Russian invasion of Ukraine, starting in February 2022, caused thousands of casualties~\cite{JerPost} and millions of refugees~\cite{UNHCR}. One of the many implications of this conflict is the impact on the Internet performance, both in Ukraine and in Russia.

In this paper we analyze the Internet performance on both sides of the Ukrainian-Russian border, showing the clear asymmetry around the Ukrainian-Russian border, with a clear performance degradation in Ukraine and a performance improvement in Russia. 
Clearly the connectivity in Ukraine was affected by infrastructure aspects such as power outages or damaged communication lines and equipment, causing routing changes and congestion along bottlenecks, and in some cases resulting in user performance degradation. In Russia, on the other hand, major content providers such as Netflix~\cite{Netflix} discontinued their service to Russian customers, and the Russian government block access to Facebook and Twitter~\cite{CBS}, resulting in content degradation.

Some preliminary insights into the impact of the war on the Internet performance in Ukraine has been discussed in~\cite{Resilience,UkrainianInternet}. The strong correlation between the Ukrainian refugee crisis and the Internet performance was introduced in~\cite{MapUkraine}.
The current work introduces a detailed comparison between the performance trend in Ukraine and in Russia during the first two months of the war based on publicly available Internet measurement data from various source, such as RIPE Atlas~\cite{RIPEatlas}, Google~\cite{Google}, Cloudflare~\cite{Cloudflare} and Speedtest~\cite{Speedtest}. 

\textbf{A note from the authors.} Our hearts are with the families of the casualties and with the refugees. We hope that our work will help in accentuating the injustice of this unnecessary conflict, and we hope that the conflict will reach a quick resolution soon.

\section{An Asymmetric Comparison}
\subsection{Internet Access Performance}
\label{AccessSec}
We analyzed the access performance in Ukraine and Russia based on data published monthly by Speedtest~\cite{Speedtest}, one of the most commonly used sites for web-based Internet performance testing. We note that Speedtest measurements are performed by connecting to local servers which are typically as close as possible to the host that initiated the test, and thus these measurement results are an indication of the \emph{local} Internet performance. 

Figure~\ref{fig:UkrSpeedLatency} provides a view at the performance impact during the first month of the war. The figure shows the performance data from the year prior to the war. 

\begin{figure}[htbp]
  \centering
  \begin{subfigure}[t]{.24\textwidth}
  \centering
  \fbox{\includegraphics[trim={4.5cm 2cm 4.5cm 2cm},clip,height=4.75\grafflecm]{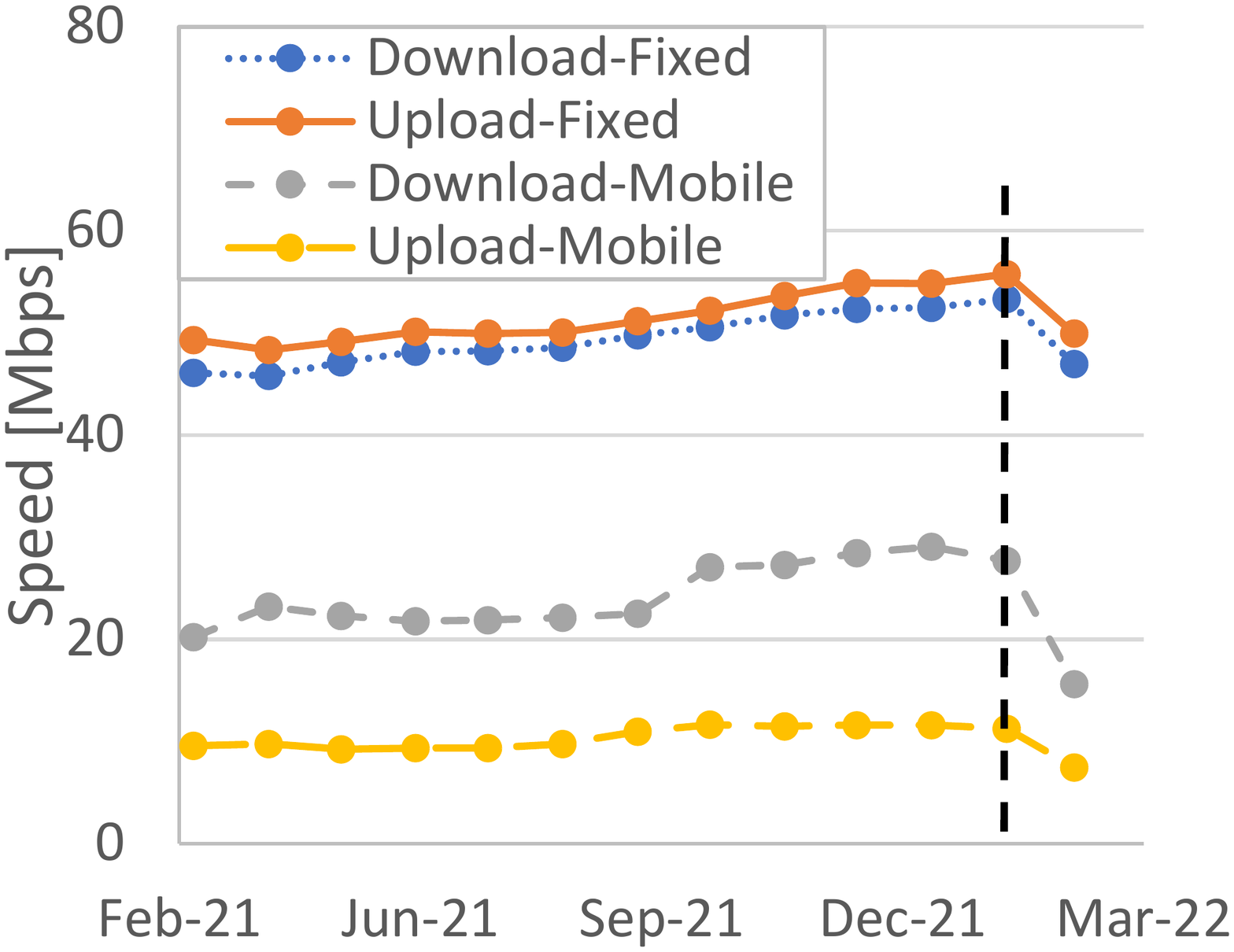}}
  \captionsetup{justification=centering}
  \caption{Ukraine: median download and upload speeds}
  \label{fig:UkrSpeed}
  \end{subfigure}%
  \begin{subfigure}[t]{.24\textwidth}
  \centering
  \fbox{\includegraphics[trim={4.5cm 2cm 4.5cm 2cm},clip,height=4.75\grafflecm]{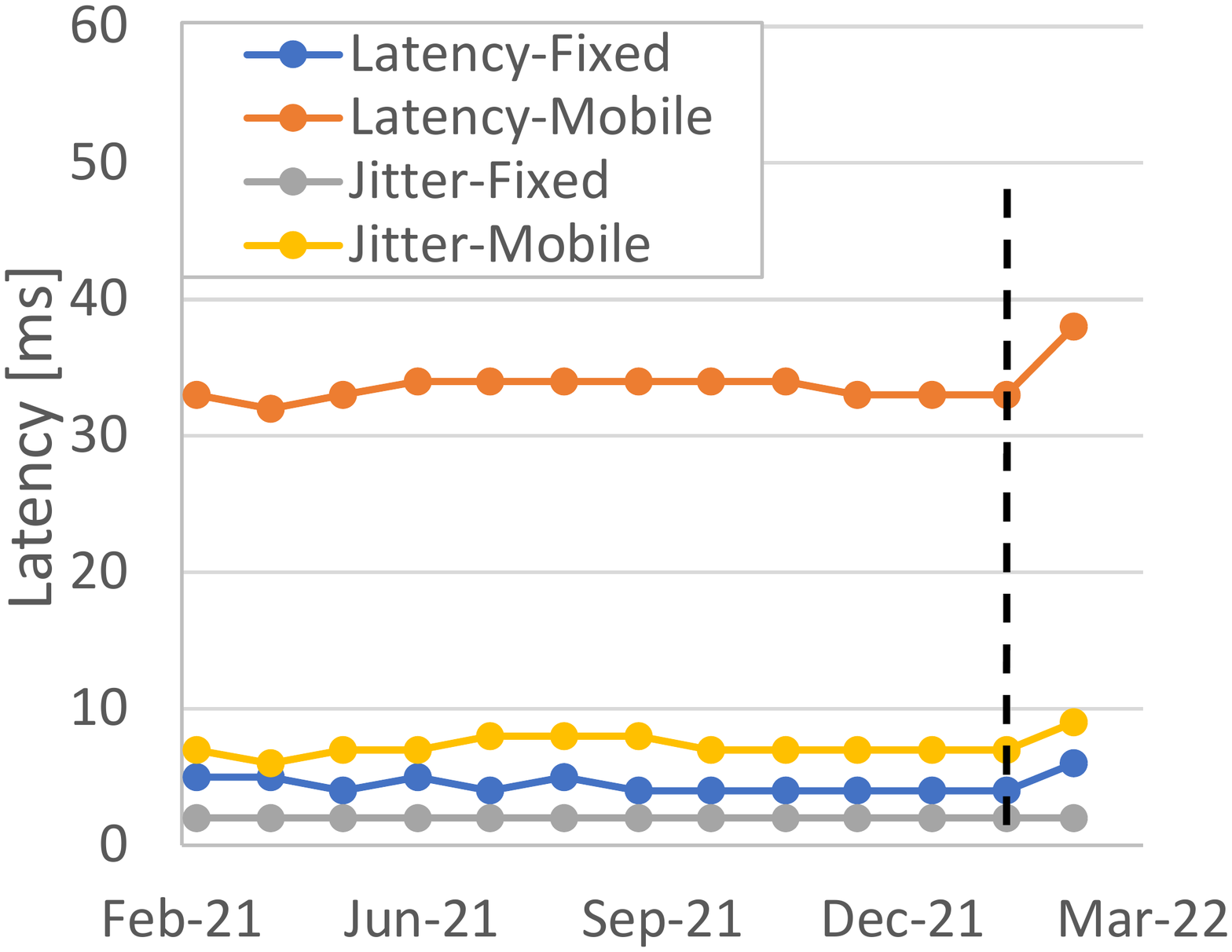}}
  \captionsetup{justification=centering}
  \caption{Ukraine: median latency and latency jitter}
  \label{fig:UkrLatency}
  \end{subfigure}%

  \begin{subfigure}[t]{.24\textwidth}
  \centering
  \fbox{\includegraphics[trim={4.5cm 2cm 4.5cm 2cm},clip,height=4.75\grafflecm]{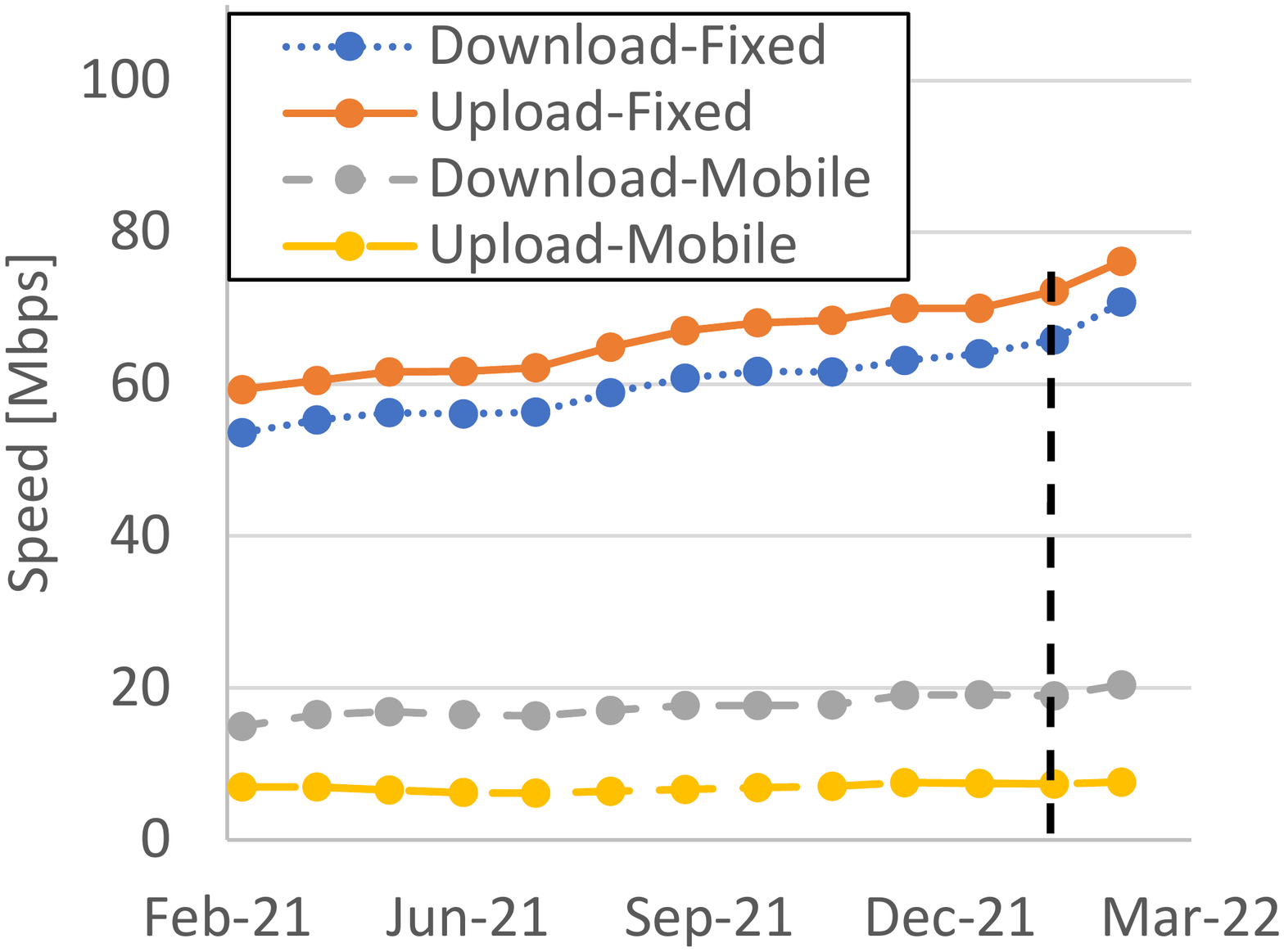}}
  \captionsetup{justification=centering}
  \caption{Russia: median download and upload speeds}
  \label{fig:RussiaSpeed}
  \end{subfigure}%
  \begin{subfigure}[t]{.24\textwidth}
  \centering
  \fbox{\includegraphics[trim={4.5cm 2cm 4.5cm 2cm},clip,height=4.75\grafflecm]{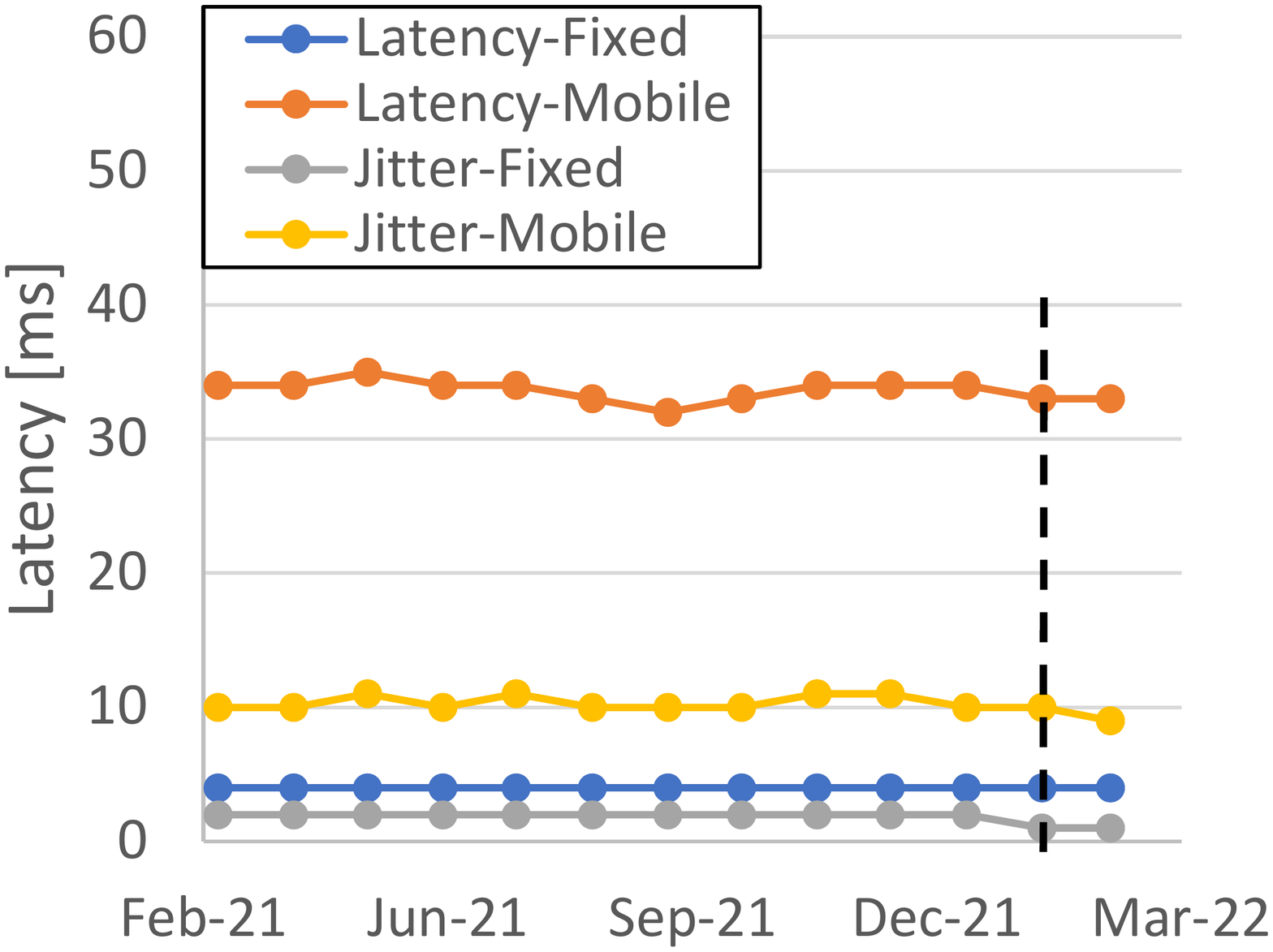}}
  \captionsetup{justification=centering}
  \caption{Russia: median latency and latency jitter.}
  \label{fig:RussiaLatency}
  \end{subfigure}%
  \caption{Speedtest results in Ukraine and in Russia. The dotted line marks the beginning of the war.}
  \label{fig:UkrSpeedLatency}
\end{figure}

\begin{figure*}[htbp]
  \centering
  \begin{subfigure}[t]{.33\textwidth}
  \centering
  \fbox{\includegraphics[trim={5cm 1cm 5cm 2cm},clip,height=6\grafflecm]{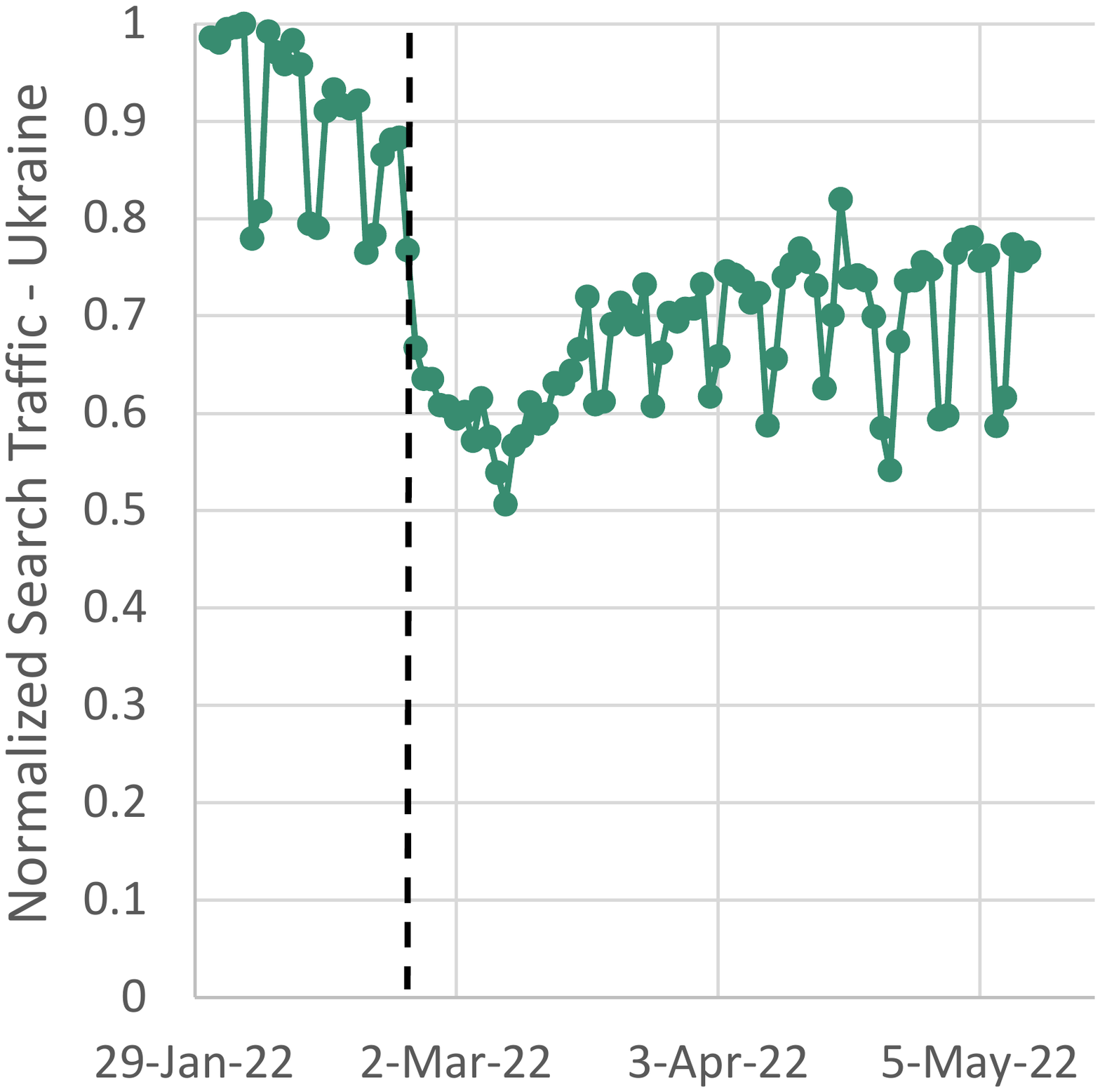}}
  \captionsetup{justification=centering}
  \caption{Normalized web search rate}
  \label{fig:UkrGoogleWeb}
  \end{subfigure}%
  \begin{subfigure}[t]{.33\textwidth}
  \centering
  \fbox{\includegraphics[trim={5cm 1cm 5cm 2cm},clip,height=6\grafflecm]{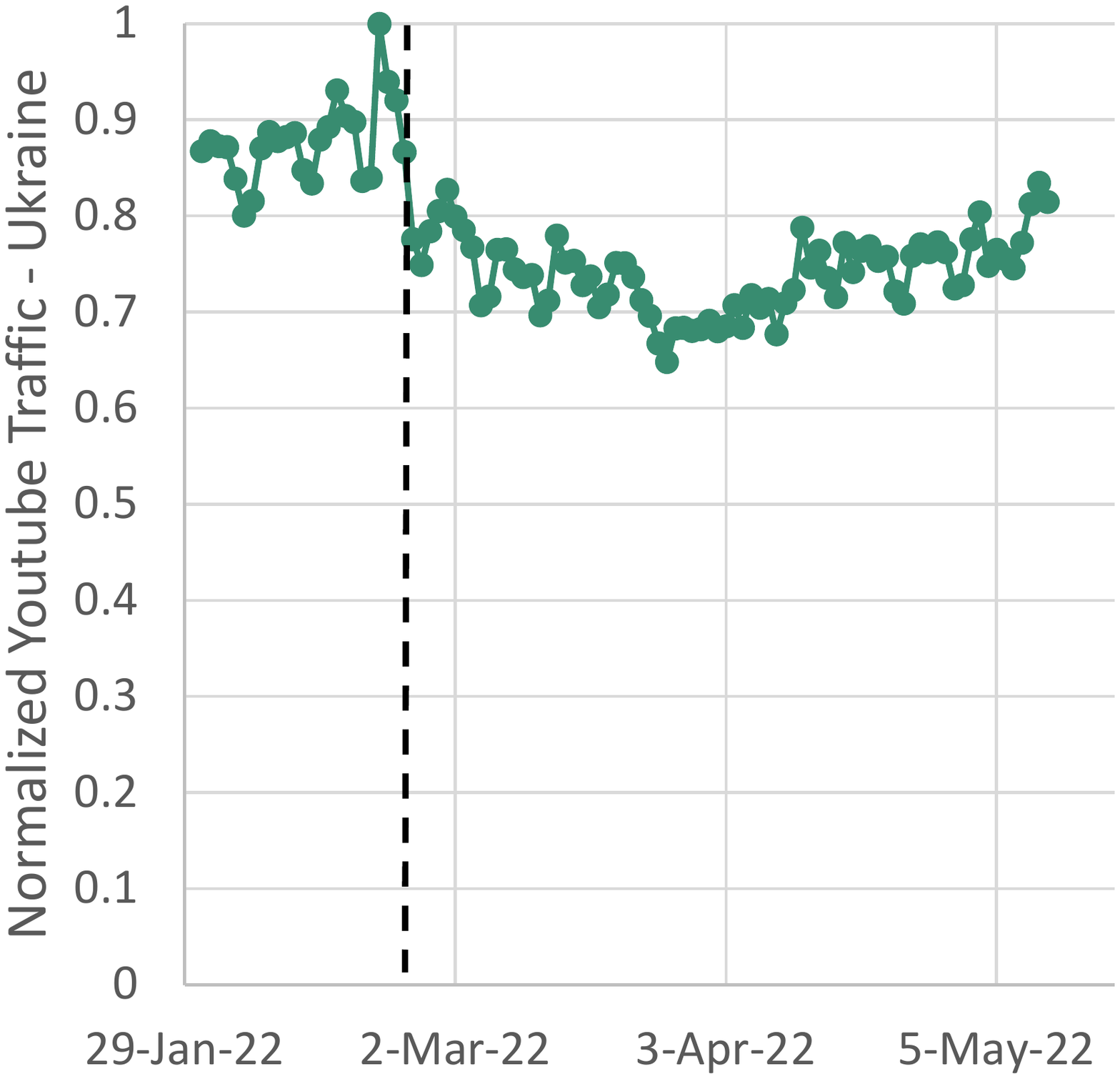}}
  \captionsetup{justification=centering}
  \caption{Normalized Youtube rate}
  \label{fig:UkrGoogleYoutube}
  \end{subfigure}%
  \begin{subfigure}[t]{.33\textwidth}
  \centering
  \fbox{\includegraphics[trim={5cm 1cm 5cm 2cm},clip,height=6\grafflecm]{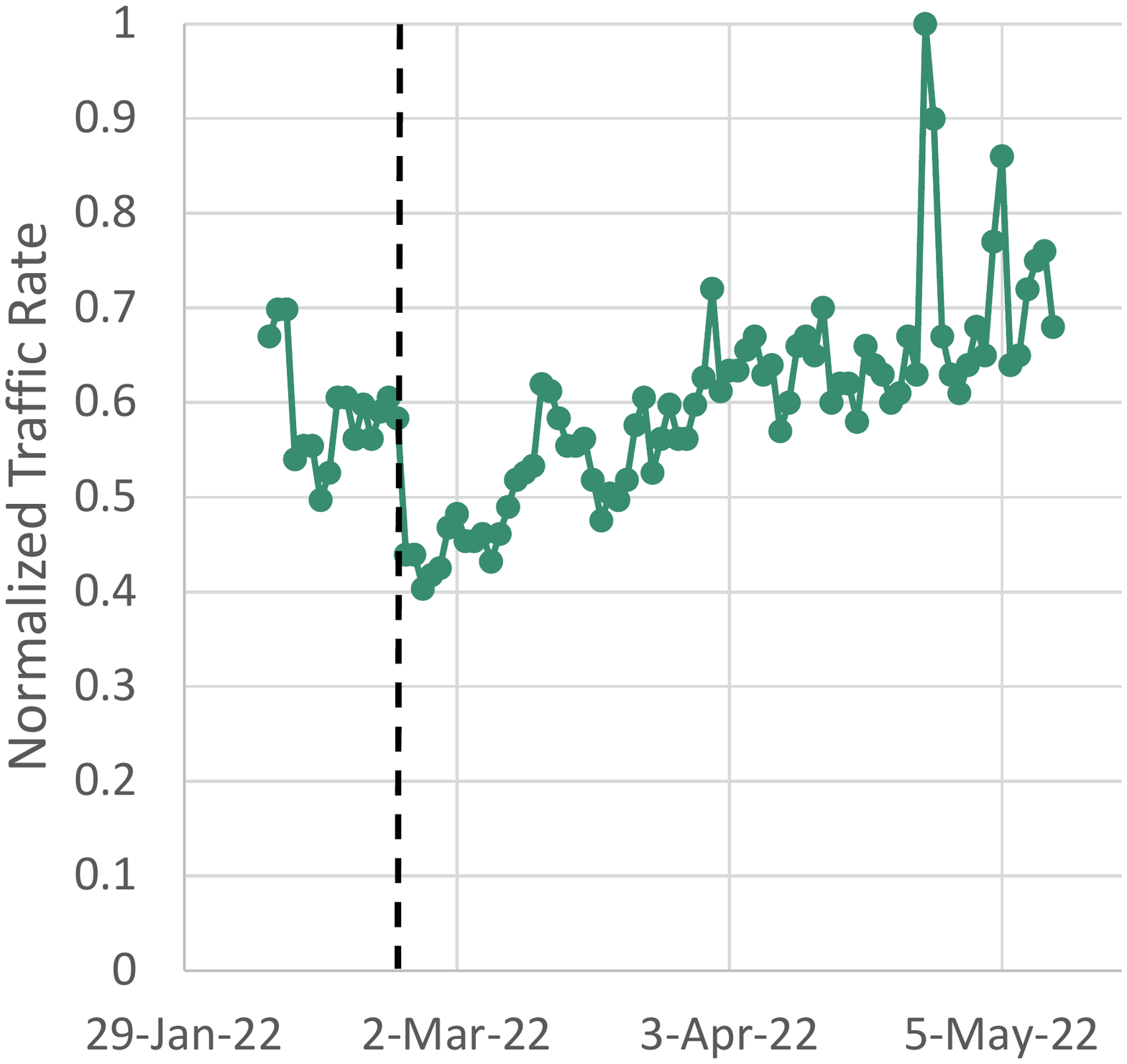}}
  \captionsetup{justification=centering}
  \caption{Normalized traffic rate}
  \label{fig:UkrCloudflare}
  \end{subfigure}%

  \begin{subfigure}[t]{.33\textwidth}
  \centering
  \fbox{\includegraphics[trim={5cm 1cm 5cm 2cm},clip,height=6\grafflecm]{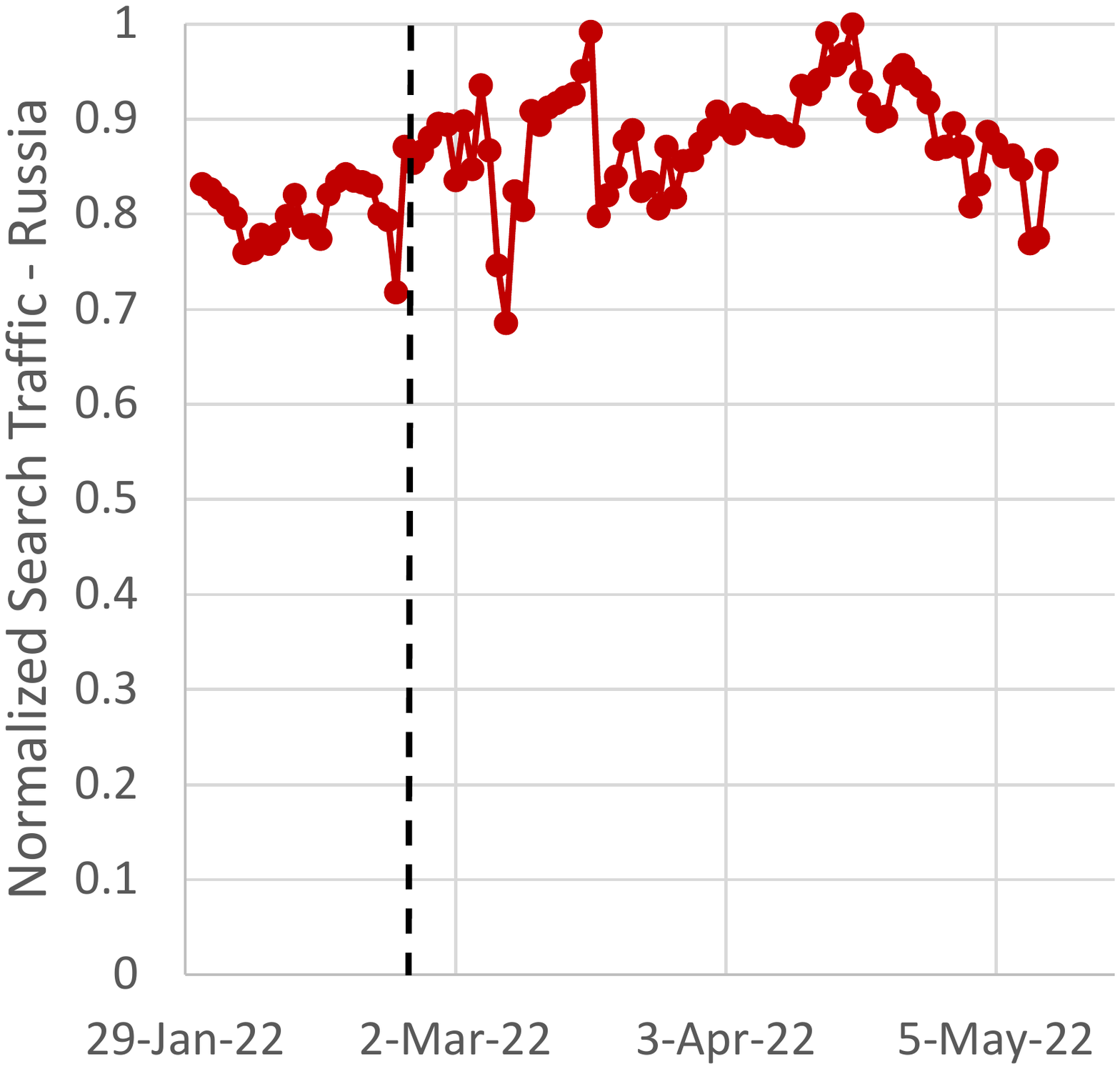}}
  \captionsetup{justification=centering}
  \caption{Normalized web search rate}
  \label{fig:RussiaGoogleWeb}
  \end{subfigure}%
  \begin{subfigure}[t]{.33\textwidth}
  \centering
  \fbox{\includegraphics[trim={5cm 1cm 5cm 2cm},clip,height=6\grafflecm]{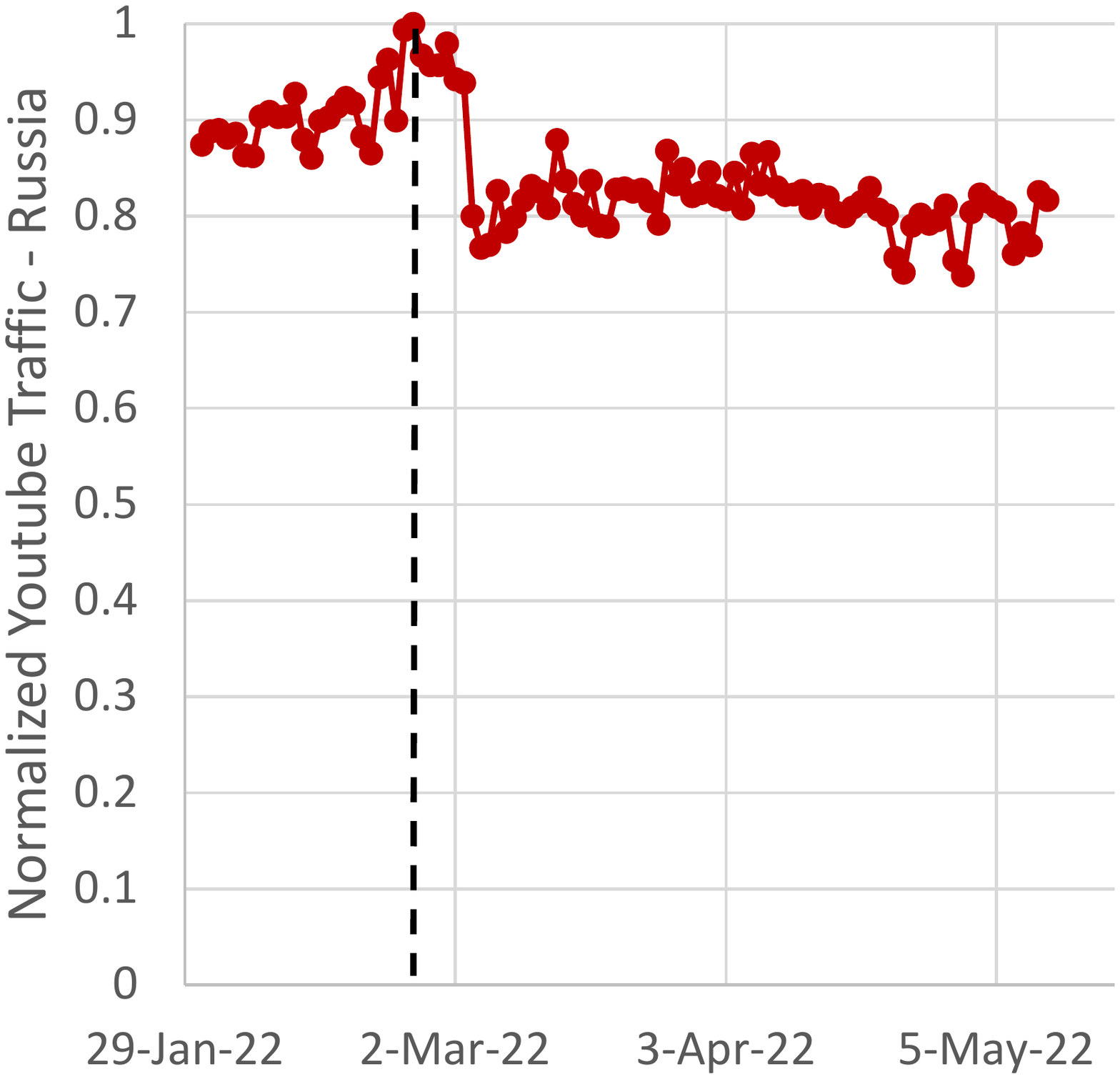}}
  \captionsetup{justification=centering}
  \caption{Normalized Youtube rate}
  \label{fig:RussiaGoogleYoutube}
  \end{subfigure}%
  \begin{subfigure}[t]{.33\textwidth}
  \centering
  \fbox{\includegraphics[trim={5cm 1cm 5cm 2cm},clip,height=6\grafflecm]{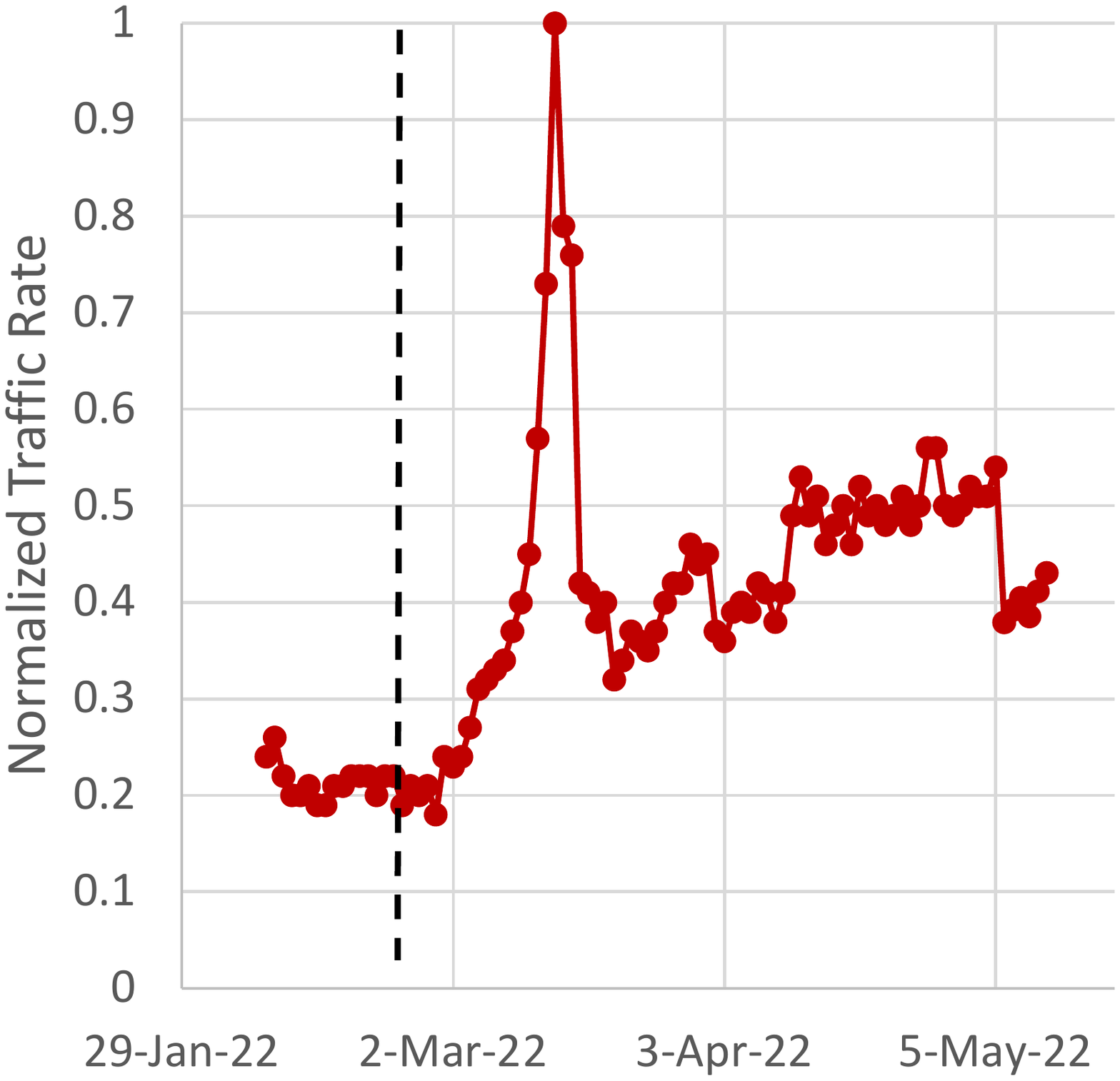}}
  \captionsetup{justification=centering}
  \caption{Normalized traffic rate}
  \label{fig:RussiaCloudflare}
  \end{subfigure}%
  \caption{Ukraine (green) vs. Russia (red) daily peak traffic rates}
  \label{fig:UkraineRussiaTraffic}
\end{figure*}

Prior to the war, i.e., in February 2022, fixed network performance in Ukraine was ranked 61st in the world, and mobile network performance was ranked 65th.  As of March 2022, the fixed network performance of Ukraine dropped to 77th, and mobile performance to 116th in the world. Over the same period of time, Russia's rank improved from 48th place to 44th place in fixed networks, and from 92nd to 90th in mobile networks.

The download and upload speed trends are illustrated in Figure~\ref{fig:UkrSpeed} and Figure~\ref{fig:RussiaSpeed}. Ukrainian upload and download speeds in fixed networks have dropped by over 10\%, while the mobile download speed has dropped by over 40\%, and the mobile upload speed degraded by over 30\%. At the same time, Russian download speeds have increased by over 7\% in both fixed and mobile networks, and upload speeds have increased by 3-5\%. The latency, as shown in Figure~\ref{fig:UkrLatency} and~\ref{fig:RussiaLatency}, was also affected significantly both in fixed and in mobile networks: the median latency in Ukraine increased by over 30\% in fixed networks and over 15\% in mobile networks, while in Russia the latency slightly improved.

These results clearly illustrate the performance degradation in Ukraine, which may be explained by infrastructure damage and outages, while the performance improvement in Russia is in line with the reduction in some of the content consumption. Note that the high latency jitter in Ukraine indicates higher network load and congestion than in previous periods.

\subsection{Traffic Rate}
We analyzed the traffic rate trends based on measurement data from by Cloudflare~\cite{Cloudflare} and Google~\cite{Google}.

A comparison of the web search traffic rate is shown in Figures~\ref{fig:UkrGoogleWeb} and~\ref{fig:RussiaGoogleWeb}. These figures illustrate the normalized daily peak rate of web search requests in each of the two countries. Similarly, Figures~\ref{fig:UkrGoogleYoutube} and~\ref{fig:RussiaGoogleYoutube} depict the normalized peak rates of Youtube traffic. Finally, Figures~\ref{fig:UkrCloudflare} and~\ref{fig:RussiaCloudflare} show the normalized peak traffic rate measured by Cloudflare.

Shortly after the beginning of the conflict there is a clear traffic rate reduction in Ukraine, shown in the three top graphs of Figure~\ref{fig:UkraineRussiaTraffic}. Comparing the first two weeks of the war to the two weeks beforehand, we found that the Google search rate in Ukraine decreased by over 30\%, and the Cloudflare rate decreased by over 20\%. Over the same period of time the Google search rate in Russia increased by 5\%, and the Cloudflare traffic rate increased by almost 25\%. This steep increase in Russia is aligned with the following explanation from Cloudflare's CEO, Matthew Prince: ``As the conflict has continued, we've seen a dramatic increase in requests from Russian networks to worldwide media, reflecting a desire by ordinary Russian citizens to see world news beyond that provided within Russia''.

The Youtube traffic rate in Ukraine showed a similar trend to the web search and the Cloudflare measurements, showing a steep decrease at the beginning of the conflict. The rate of Youtube access in Russia was reduced by over 10\% at the beginning of March, which is aligned with the Russian government's announcement about blocking Facebook and Twitter~\cite{CBS}, thus reducing the exposure to many of the Youtube links that were conventionally accessed from these social networks prior to the blocking decision.


\subsection{RTT and Loss}
One of the metrics that significantly degraded as of the beginning of the war was the median latency, as previously shown in Figure~\ref{fig:UkrSpeedLatency}.
We now show that a similar latency trend can be seen in the RIPE Atlas Round-Trip Time (RTT) measurements we analyzed, shown in Figure~\ref{fig:RTT}. We analyzed RTT measurements of about 200 probes in Ukraine and about 600 probes in Russia over a period of more than 3 months, using the built-in periodic Ping measurements to the K-Root DNS servers, summing up to over 20 million RTT measurements. The number of connected probes was analyzed on a per-day basis. Data was extracted from RIPE Atlas using the Magellan tool~\cite{RIPEmagellan}

The median RTT is illustrated in Figure~\ref{fig:RTT}. The figure illustrates an increased RTT in Ukraine starting on the first week of the war, while the RTT in Russia has very slight fluctuations over the observed period. The measured packet loss rate over these measurements is illustrated in Figure~\ref{fig:Loss}, showing a large loss rate increase in Ukraine, while the loss rate in Russia had very slight variability over most of the observed period.

\begin{figure}[htbp]
  \centering
  \begin{subfigure}[t]{.24\textwidth}
  \centering
  \fbox{\includegraphics[trim={5cm 2cm 5cm 2cm},clip,height=5.3\grafflecm]{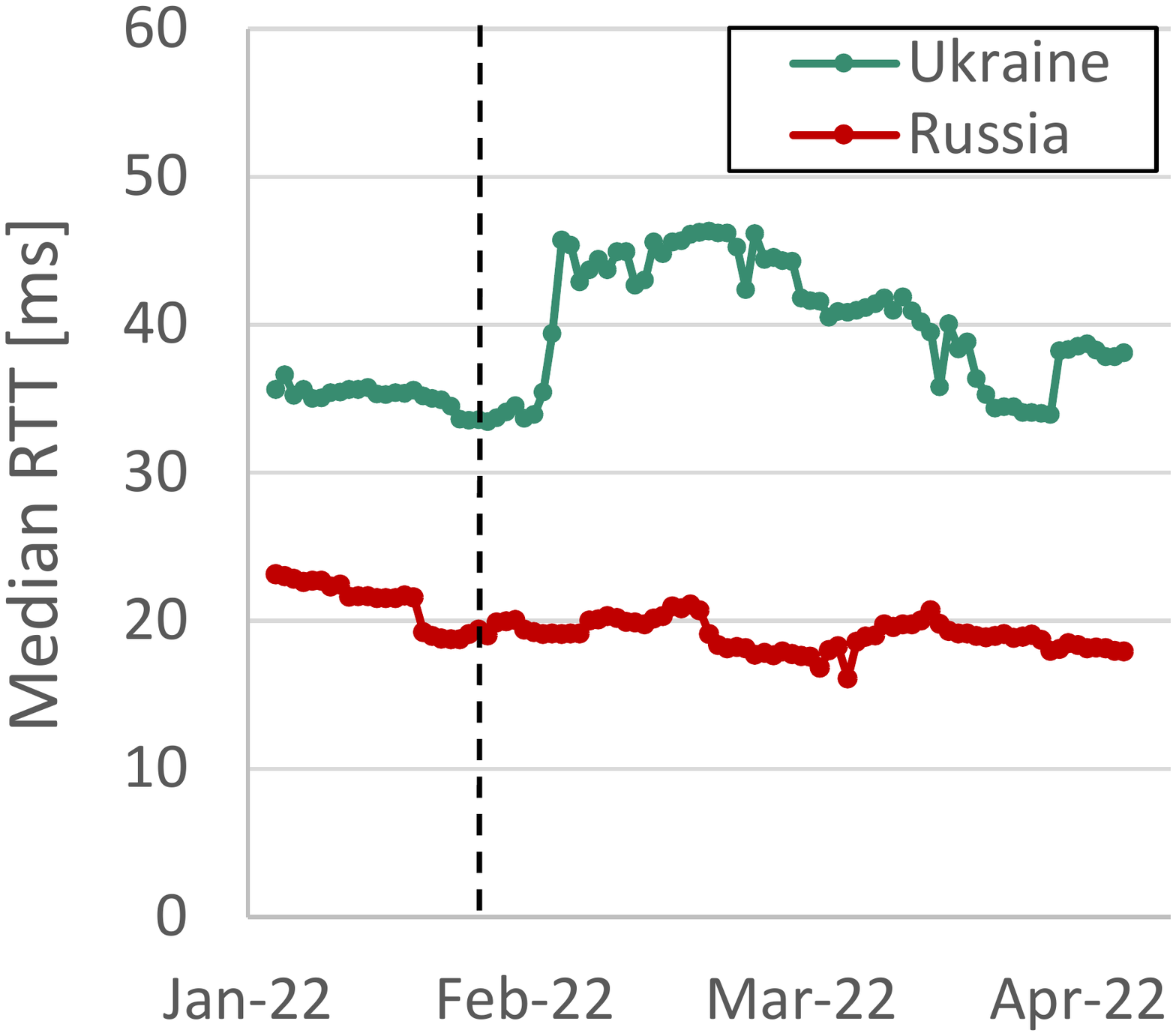}}
  \captionsetup{justification=centering}
  \caption{Median RTT}
  \label{fig:RTT}
  \end{subfigure}%
  \begin{subfigure}[t]{.24\textwidth}
  \centering
  \fbox{\includegraphics[trim={5cm 2cm 5cm 2cm},clip,height=5.3\grafflecm]{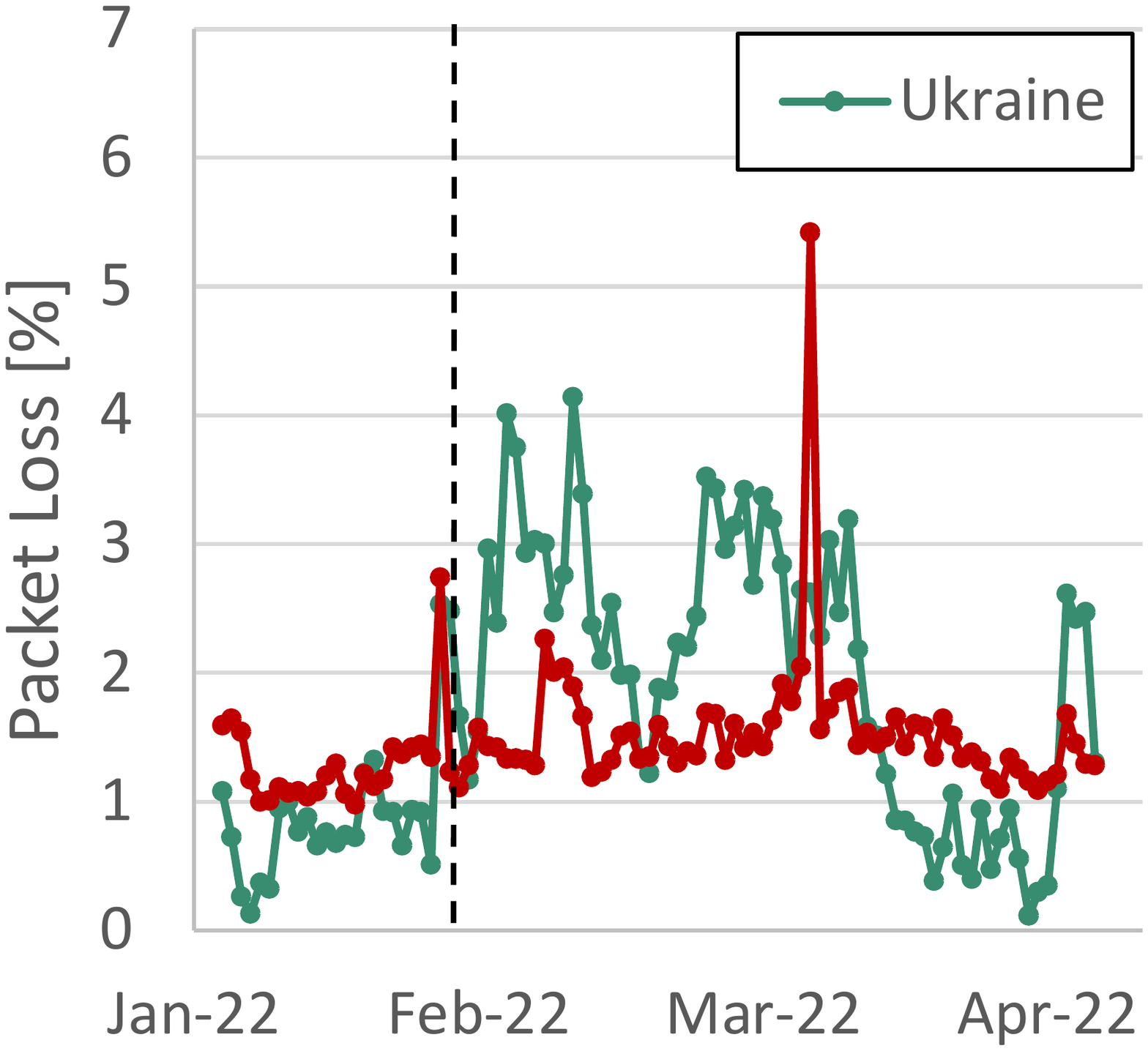}}
  \captionsetup{justification=centering}
  \caption{Loss percentage}
  \label{fig:Loss}
  \end{subfigure}%
  \caption{RIPE Atlas RTT measurements from Ukraine probes to K-Root servers}
  \label{fig:RTTandLoss}
\end{figure}

We further investigated the RTT changes in Ukraine by following the RIPE Atlas Traceroute tests to the K-Root servers. We compared the traces in 21 February to the traces in 21 March, and found that for many of the RIPE Atlas probes the path to the K-Root servers have changed during this month. Specifically, many of the probes which used a routed through Germany in February switched to a different path through the Netherlands in March. This routing change was a common change in many of the traces during this period, and can be explained by routing constraints caused by infrastructural aspects.

The RTT and loss rate measurements from the RIPE Atlas platform confirm the general trend of the performance degradation in Ukraine, as opposed to the performance in Russia. This trend is shown to be visible not only in local traffic, as shown in Section~\ref{AccessSec}, but also over a large distance, as shown in the RIPE Atlas measurement analysis.

\section{Conclusion}
\label{ConcSec}
We presented a measurement study of the Internet performance in Ukraine and in Russia during the first two months after the beginning of the conflict on February 2022. This analysis presents an asymmetric picture: Internet performance in Ukraine significantly degraded, while the performance in Russia improved. Ukraine suffered from various infrastructure issues and from the refugee crisis, resulting in reduced download and upload speeds, increased latencies, and high loss rate. Russian users, on the other hand, experienced improved download and upload speeds, likely due to reduced streaming consumption from leading services such as Netflix and Youtube.

\ifdefined\TechReport
\else
\section*{Appendix: Ethics}
\label{EthicsSec}
This work does not raise any ethical issues. Specifically, any ethical issues that are related to the war in Ukraine are outside the scope of this work. From a privacy perspective, all the data that is analyzed in this work is publicly available, and the analysis produces statistics and trends, without compromising any privacy aspects.
\fi

\bibliographystyle{abbrv}
\bibliography{Uk}

\end{document}